\documentclass[aps,showpacs,twocolumn,superscriptaddress]{revtex4}

\newcommand{\bea}{\begin{eqnarray}}
\newcommand{\eea}{\end{eqnarray}}
\newcommand{\beq}{\begin{equation}}
\newcommand{\eeq}{\end{equation}}

\usepackage{amsmath}
\usepackage{amssymb}
\usepackage{latexsym}
\usepackage{graphicx}

\usepackage{graphics}
\usepackage{psfig}
\usepackage{epsfig}
\usepackage{color}
\usepackage{changebar}

\begin{document}


\title{Slice Stretching at the Event Horizon \hbox{when Geodesically Slicing the Schwarzschild Spacetime with Excision}}


\author{Bernd Reimann}
\affiliation{Max Planck Institut f\"ur Gravitationsphysik,
Albert Einstein Institut, Am M\"uhlenberg 1, 14476 Golm, Germany}
\affiliation{Instituto de Ciencias Nucleares, Universidad Nacional Aut{\'o}noma de M{\'e}xico, A.P. 70-543, M{\'e}xico D.F. 04510, M{\'e}xico}


\date{August 25, 2004}


\begin{abstract}
Slice-stretching effects are discussed as they arise at the event horizon when geodesically slicing the extended Schwarzschild black-hole spacetime while using singularity excision.
In particular, for Novikov and isotropic spatial coordinates the outward movement of the event horizon (``slice sucking'') and the unbounded growth there of the radial metric component (``slice wrapping'') are analyzed.
For the overall slice stretching, very similar late time behavior is found when comparing with maximal slicing.
Thus, the intuitive argument that attributes slice stretching to singularity avoidance is incorrect.
\end{abstract}


\pacs{
04.20.Jb,   
04.25.Dm,   
04.70.Bw,   
95.30.Sf    
\quad Preprint numbers: AEI-2004-039
}


\maketitle


\section{Introduction}
\label{sec:introduction}
When evolving a spacetime containing a physical singularity without making use of a shift, the foliation usually is of pathological nature as so-called ``slice-stretching'' effects develop \cite{Centrella86}.
Here ``slice sucking'' arises in the form of outward-drifting coordinate locations as the corresponding observers are falling towards the singularity.
This infall takes place in a differential manner, leading to large proper distances in between neighboring observers with ``slice wrapping'' showing up in the form of large gradients in the radial metric function.

In the past, slice-stretching effects have been often attributed to the singularity-avoiding behavior of a foliation by the following intuitive argument: 
For such slices the lapse collapses to zero in the strongly curved ``interior'' region and the evolution of the metric essentially freezes there, while it marches ahead further ``out'' in order to evolve a large fraction of the spacetime.
Hence, singularity-avoiding slices ``wrap up around the singularity'' \cite{Anninos95_2} which causes ``large amounts of shear in the coordinate grid'' \cite{Bernstein89}.

To study whether the singularity-avoiding property of a foliation plays a role for the overall slice stretching, I will in this paper for the extended Schwarzschild spacetime compare slice-stretching effects for geodesic and maximal slicing.

Characterized by unit lapse and vanishing shift, geodesic slicing represents the ``simplest'' gauge choice which, however, does not avoid physical singularities.
In particular, starting with the time-symmetric conformally flat Einstein-Rosen bridge \cite{Einstein35} of the extended Schwarzschild spacetime, an observer at the throat is initially at the event horizon \hbox{$r_{EH} = 2M$} and falls freely into the singularity. 
Hence, if the singularity is part of the grid, a numerical simulation faces infinite curvature and has to crash after evolving for the free-fall time given by $\pi M$.
This fact has been used frequently in numerical relativity when performing the so-called ``crash test'' as e.g. in \cite{Bernstein89} or \cite{Anninos95}.
In addition, when not excising the singularity, the analytically known non-trivial evolution of the \hbox{3-metric} can be used up to the time $\pi M$ for testing numerical codes and for studying the numerical behavior of different formulations of the Einstein evolution equations, see e.g.\ \cite{Imbiriba2004} and \cite{Jansen2003}.

In the following, however, I am only interested in the portion of the geodesic slices lying in the exterior parts of the spacetime.
From the viewpoint of numerical relativity, I make use of a code capable of excising the interior region in between the ``left-hand'' and ``right-hand'' event horizon.
For Novikov spatial coordinates this idea is shown in Fig.~1 of \cite{Bruegmann96}.
In the same reference it has also been demonstrated numerically that geodesic slicing together with singularity excision can be used to evolve a single black hole for considerably more than $\pi M$.

When studying the late time behavior of the geodesic slices, I will for simplicity concentrate on the event horizon acting as a ``marker'' for slice-stretching effects.
In particular, the location of the event horizon in terms of Novikov and isotropic spatial coordinates will be determined as a function of time together with the behavior of the radial component of the 3-metric there. 

These results will then be compared both analytically and numerically with statements obtained for maximal slicing in \cite{mythesis,mypaper1,mypaper2,mypaper3}.
Being motivated geometrically and characterized by the condition that the trace of the extrinsic curvature vanishes at all times \cite{York79}, maximal slices avoid the Schwarzschild singularity by approaching the limiting slice \hbox{$r = 3M/2$} asymptotically \cite{Estabrook73}.
The main result of this comparison is that the overall slice stretching at the event horizon for both geodesic and maximal slicing in leading order is found to be proportional to time.
Hence the intuitive argument that attributes slice stretching to singularity avoidance turns out to be incorrect.   

The paper is organized as follows:
In Sec.~\ref{sec:slicestretching} slice-stretching effects at the event horizon are discussed, studying in Subsec.~\ref{subsec:radial} the cycloidal motion of radial geodesics and focusing in \hbox{Subsecs.~\ref{subsec:Novikov} and \ref{subsec:isotropic}} on Novikov and isotropic spatial coordinates, respectively.
I conclude in Sec.~\ref{sec:conclusion}.


\section{Slice Stretching at the Event Horizon}
\label{sec:slicestretching}


\subsection{Cycloidal motion of radial geodesics}
\label{subsec:radial}
Geodesic slicing is characterized by unit lapse and vanishing shift which define Gaussian normal coordinates.
Those are comoving in the sense that radially freely falling observers are at rest and the time coordinate measures proper time.
One important property of Gaussian normal coordinates is that the geodesics defining the coordinates remain orthogonal to all constant time hypersurfaces, and transformations between different spatial coordinates are hence time-independent. 

In particular, referring to Schwarzschild coordinates, a radial geodesic starting at the singularity at \hbox{$r = 0$} performs a cycloidal motion out to some maximal radius $\tilde{r}$ and back as pointed out in more detail in \S 25.5 and \S 31.3 of \cite{Misner73}.
In terms of \hbox{$\tilde{r} > 2M$}, the geometry of the Schwarzschild spacetime is described by the line element
\beq
\label{eq:rmaxmetric}
	ds^{2} = - d\tau^2
                 + \frac{1}{1 - 2M/\tilde{r}}
                   \left( \frac{\partial r}{\partial \tilde{r}} \right)^2 
			\:d\tilde{r}^{2}
                 + r^2\:d\Omega^2.
\eeq
Here by integrating the geodesic equation one can see that \hbox{$r = r(\tau,\tilde{r})$} is given implicitly by
\beq
\label{eq:tau}
	\tau = \tilde{r} \left[
     	       \sqrt{\frac{r}{2M} \left( 1 - \frac{r}{\tilde{r}} \right)}
             + \sqrt{\frac{\tilde{r}}{2M}} 
               \ \begin{rm}{arccos}\end{rm}
	       \sqrt{\frac{r}{\tilde{r}}} \right].   
\eeq
Furthermore, following \cite{Bruegmann96}, it turns out that by implicit differentiation 
\beq
\label{eq:diff}
	\frac{\partial r}{\partial \tilde{r}} = \frac{3}{2} \left[
				  1 - \frac{r}{3\tilde{r}} 
				  + \sqrt{\frac{\tilde{r}}{r} - 1} 
		   	          \ \begin{rm}{arccos}\end{rm} 
				  \sqrt{\frac{r}{\tilde{r}}} \right] 
\eeq
is found (but note a missing root in the formula given in that reference).

Concentrating for a discussion of slice stretching on the event horizon, \hbox{$r = r_{EH} = 2M$}, in leading order (to be denoted by $\simeq$) from (\ref{eq:tau}) one can infer
\beq
\label{eq:rmaxmove}
	\tilde{r}_{EH}
	\simeq \frac{2 M^{\frac{1}{3}}}{\pi^{\frac{2}{3}}} \tau^{\frac{2}{3}}.
\eeq
Interested in the behavior of the radial component $G_{\tilde{r}\tilde{r}}$ of the line element (\ref{eq:rmaxmetric}) there, when using (\ref{eq:diff}) at late times
\beq
\label{eq:rmaxG}
	\left. G_{\tilde{r}\tilde{r}} \right|_{\tilde{r}_{EH}} 
	\simeq \frac{9\pi^{\frac{4}{3}}}{16M^{\frac{2}{3}}} \tau^{\frac{2}{3}} 
\eeq
is found.

In a next step it is then desirable to combine the coordinate-dependent effects of slice sucking and slice wrapping, (\ref{eq:rmaxmove}) and (\ref{eq:rmaxG}), into one coordinate-independent quantity measuring the overall slice stretching. 
Here I want to point out that for maximal slicing in \cite{mypaper3} the overall slice stretching has been characterized in a global sense by integrals of metric quantities such as the square root of the radial metric component.
Those integrations in space have been carried out over the throat and, when integrating e.g.\ from the left- to the right-hand event horizon, for ``numerically favorable'' boundary conditions a divergence being in leading order proportional to time has been found, see \cite{mypaper3} for details.
Note, however, that one can not carry out such an integration over the throat for slicings like geodesic slicing which do not avoid but hit the singularity.

Therefore I now want to characterize the overall slice stretching in a local sense only.
For this task I introduce the corresponding indefinite integrals and evaluate them at the right-hand event horizon setting \hbox{$r = r_{EH} = 2M$} and using previously derived results.
Integrating the square root of the radial metric component in this sense, 
\bea
\label{eq:leadinggeodesic}
	 & & \int^{\tilde{r}_{EH}} \frac{3 \left[
				  1 - \frac{r_{EH}}{3y}
				  + \sqrt{\frac{y}{r_{EH}} - 1} 
		   	          \ \begin{rm}{arccos}\end{rm} 
				  \sqrt{\frac{r_{EH}}{y}} \right]}
				    {2 \sqrt{1 - 2M/y}} \: dy \nonumber \\
	& & \ \ \ \ \ \ \ \ \ \ \ \ \ 
		\simeq \frac{\pi}{2^{\frac{3}{2}} M^{\frac{1}{2}}} \tilde{r}_{EH}^{\frac{3}{2}}
		\simeq \tau,
\eea
for geodesic slicing one obtains a divergence which is in leading order linear in time.

For maximal slicing one can write the radial metric component as \hbox{$G_{xx} = x^4 \Psi^{12}(x) / r^4$}.
This formula was found in \cite{mypaper1} for evolutions using isotropic grid coordinates $x$ where the conformal factor is given by \hbox{$\Psi(x) = 1 + M/2x$}.
In terms of these coordinates the location of the right-hand event horizon is described by  
\beq
	x_{EH}^+ \simeq \frac{3^{\frac{5}{6}} M^{\frac{2}{3}}}{2^{\frac{2}{3}}} \tau^{\frac{1}{3}}
\eeq
as pointed out in \cite{mypaper3}, and hence with
\beq
\label{eq:leadingmaximal}
	\int^{x_{EH}^+} \frac{y^2 \Psi^6(y)}{r_{EH}^2} \: dy 
	\simeq \frac{x_{EH}^{+ \: 3}}{3 r_{EH}^2}
	\simeq \frac{3 \sqrt{3}}{16} \tau	
\eeq
also for maximal slicing a behavior of order \hbox{${\cal O}(\tau)$} is obtained \cite{commentprefactor}.

The coordinate-independent observations (\ref{eq:leadinggeodesic}) and (\ref{eq:leadingmaximal}) clearly are in contradiction to the argument that slice stretching is due to singularity avoidance.
In order to make further statements regarding the ``splitting'' of the overall slice stretching into slice sucking and slice wrapping, I will discuss in the next subsection geodesic slicing in terms of (more familiar) Novikov coordinates.


\subsection{Novikov coordinates}
\label{subsec:Novikov}
The Novikov spatial coordinate $R^*$ \cite{Novikov62} is related time-independently to $\tilde{r}$ by
\beq
\label{eq:rmaxRstar}
	\tilde{r} = 2M (R^{*2} + 1)
\eeq
where the isometry \hbox{$R^* \longleftrightarrow - R^*$} is mapping the two ``universes'' of the extended Schwarzschild spacetime into each other.
Considering (\ref{eq:rmaxmove}), one can then observe that in terms of $R^*$ slice stretching takes place in a symmetric manner as the locations of the left- and right-hand event horizon at late times are given by
\beq
	R_{EH}^{*\pm} \simeq \pm \frac{1}{(\pi M)^{\frac{1}{3}}} \tau^{\frac{1}{3}}.
\eeq
For the radial component of the metric, 
\beq
	G_{R^*R^*} (\tau,R^*) =  16M^2 (R^{*2} + 1) 
                   \left( \frac{\partial r}{\partial \tilde{r}} \right)^2,
\eeq
using (\ref{eq:diff}) in leading order at both left- and right-hand event horizon an unbounded growth like
\beq
	\left. G_{R^*R^*} \right|_{R_{EH}^{*\pm}} 
	\simeq 9 (\pi M)^{\frac{2}{3}} \tau^{\frac{4}{3}} 
\eeq
is found.

I want to emphasize here that slice sucking and slice wrapping at the event horizon - for Novikov spatial coordinates - takes place in order \hbox{${\cal O}(\tau^{1/3})$} and \hbox{${\cal O}(\tau^{4/3})$}, respectively.
In \cite{mypaper2} the same late time behavior - but for isotropic grid coordinates - has been found for maximal slicing (at the right-hand event horizon in the context of even or ``zgp'' boundary conditions).

In order to compare the slice-stretching effects for the same choice of spatial coordinates, in the next subsection I will geodesically slice puncture data.


\subsection{Isotropic coordinates}
\label{subsec:isotropic}
Isotropic spatial coordinates can be introduced again in a time-independent way by
\beq
\label{eq:rmaxx}
	\tilde{r} = x \Psi^2(x)
\eeq
making use of the conformal factor \hbox{$\Psi(x) = 1 + M/2x$}.
Since the isometry \hbox{$x \longleftrightarrow M^2/4x$} is present, with (\ref{eq:rmaxmove}) and (\ref{eq:rmaxx}) one may readily verify 
\beq
	x_{EH}^{+} = \frac{M^2}{4x_{EH}^{-}} \simeq \frac{2M^{\frac{1}{3}}}{\pi^{\frac{2}{3}}} \tau^{\frac{2}{3}}
\eeq
and observe that in leading order the location of the right-hand event horizon grows in order \hbox{${\cal O}(\tau^{2/3})$} whereas the left-hand event horizon approaches the puncture like \hbox{${\cal O}(\tau^{-2/3})$}.
Furthermore, it turns out that the radial component of the 3-metric - often rescaled by $\Psi^4(x)$ to focus on the dynamical features in the metric rather than on the static singularity at \hbox{$x = 0$} - is given by
\beq
\label{eq:rescaledgxx}
        g_{xx} (\tau,x) =  \frac{G_{xx}(\tau,x)}{\Psi^4(x)} 
                        =  \left( \frac{\partial r}{\partial \tilde{r}} \right)^2.
\eeq
When using (\ref{eq:diff}) at the left- and right-hand event horizon for $g_{xx}$ an unbounded growth of order \hbox{${\cal O}(\tau^{2/3})$} is found,
\beq
	\left. g_{xx} \right|_{	x_{EH}^\pm} 
	\simeq \frac{9\pi^{\frac{4}{3}}}{16M^{\frac{2}{3}}} \tau^{\frac{2}{3}}, 
\eeq
whereas when including the conformal factor for $G_{xx}$ a behavior of order \hbox{${\cal O}(\tau^{10/3})$} and \hbox{${\cal O}(\tau^{2/3})$} is obtained,
\beq
	\left. G_{xx} \right|_{x_{EH}^-}  
	\simeq \frac{144}{\pi^{\frac{4}{3}}M^{\frac{10}{3}}} \tau^{\frac{10}{3}} 
	\ \ \ \begin{rm}{and}\end{rm} \ \ \ 
	\left. G_{xx} \right|_{x_{EH}^+}  
	\simeq \frac{9\pi^{\frac{4}{3}}}{16M^{\frac{2}{3}}} \tau^{\frac{2}{3}}. 
\eeq

\begin{figure} [!ht]
	\noindent
	\epsfxsize=80mm \epsfysize=52.5mm \epsfbox{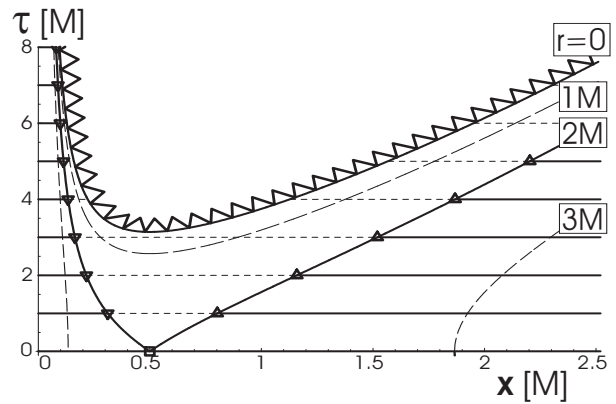} 
	\caption{The geodesically sliced Schwarzschild spacetime in isotropic coordinates is shown, denoting the location of the left- and right-hand event horizon by downward- and upward-pointing triangles, respectively. Note that in numerical practice one would usually excise from the puncture at \hbox{$x = 0$} up to a ``ghost zone'' to the left of the right-hand event horizon.}
	\label{fig:Novikov}
\end{figure}

The geodesic slicing of puncture data has been investigated numerically in \cite{Bruegmann96}.
Here I show the corresponding spacetime diagram in Fig.~\ref{fig:Novikov}.
\begin{figure} [!ht]
	\noindent
	\epsfxsize=80mm \epsfysize=102.5mm \epsfbox{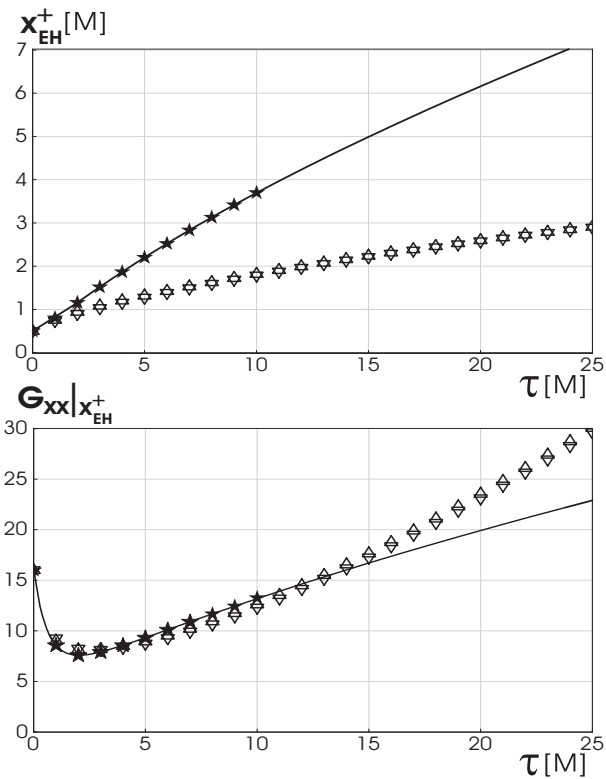} 
	\caption{Slice sucking and slice wrapping are shown as taking place for isotropic coordinates at the event horizon. Here the numerically obtained data for geodesic slicing (shown as stars) is in excellent agreement with the analytically predicted line, with both $x_{EH}^+$ and \hbox{$G_{xx}|_{x_{EH}^+}$} being of order ${\cal O}(\tau^{2/3})$. \hbox{For maximal slicing} these slice-stretching effects are of order ${\cal O}(\tau^{1/3})$ and ${\cal O}(\tau^{4/3})$, obtaining numerically very similar results when demanding even or ``zgp'' boundary conditions (denoted by upward- or downward-pointing triangles, respectively).}
	\label{fig:isotropicSS}
\end{figure}

Furthermore, by making use of the regularized spherically symmetric code described in \cite{Alcubierre04RegCode}, I have evolved Schwarzschild puncture data numerically with geodesic slicing and, for comparison, maximal slicing.
For simulations using $30,000$ grid points and a resolution of \hbox{$\triangle x = 0.001M$} the location of the right-hand event horizon and the value of the radial metric component there, $x_{EH}^+$ and \hbox{$G_{xx} |_{x_{EH}^+}$}, are shown as a function of time in Fig.~\ref{fig:isotropicSS}.

One should observe in this figure that for geodesic slicing the numerically found slice sucking and slice wrapping is up to \hbox{$\tau = 10M$} in excellent agreement with the analytically predicted results when excising from the puncture up to $0.98$ times the value of $x_{EH}^{+}$.
Due to growing errors at the excision boundary, however, the run fails shortly afterwards.

Foliations of Schwarzschild puncture data using maximal slicing, its lapse arising from the elliptic equation \hbox{$\triangle \alpha = R \alpha$}, have been studied both analytically and numerically in \cite{mythesis,mypaper1,mypaper2,mypaper3}. 
In particular, for runs demanding symmetry with respect to the throat or for the puncture evolution imposing a vanishing gradient of the lapse at the puncture, i.e.\ for even or ``zgp'' boundary conditions, slice sucking and slice wrapping at the right-hand event horizon have been found to be of order \hbox{${\cal O}(\tau^{1/3})$} and \hbox{${\cal O}(\tau^{4/3})$}, respectively.
Numerical results are shown here in Fig.~\ref{fig:isotropicSS}.


\section{Conclusions}
\label{sec:conclusion}
I have studied the slice-stretching effects which are present when geodesically slicing the extended Schwarzschild spacetime while making use of singularity excision.
The analysis has been carried out at the event horizon in terms of Novikov and isotropic spatial coordinates.

Independent of the coordinate choice, the overall slice stretching has been found to be proportional to time and hence to be comparable to the one arising for maximal slicing.
Its splitting into slice sucking and slice wrapping, however, for the same choice of spatial coordinates turned out to be different.

The intuitive argument that attributes slice stretching to singularity avoidance has been found to be incorrect.
Instead, for evolutions with vanishing shift, slice sucking and slice wrapping are caused by the differential infall of Eulerian observers.

Taking the maximally sliced Schwarzschild spacetime as an example, it is furthermore of interest to study analytically whether these effects can be avoided by making use of a geometrically motivated shift.
I will report on work in this direction in a further paper \cite{mypaper5}.


\bigskip
\acknowledgments
It is a pleasure for me to thank M.~Alcubierre, B.~Br\"ugmann and D.~Pollney for helpful comments on the manuscript. 
Furthermore, I am grateful to \hbox{J.A.}~Gonz{\'a}lez for making use of his excision routine. 


\bibliographystyle{apsrev}

\bibliography{myreferences}

\begin{thebibliography}{19}
\bibitem[{com()}]{commentprefactor}
\bibinfo{note}{I want to point out that the late time behavior of the
  indefinite integrals (\ref{eq:leadinggeodesic}) and (\ref{eq:leadingmaximal})
  - measuring for geodesic and maximal slicing the overall slice stretching at
  the right-hand event horizon - seems to be related closely to the value of
  the lapse there. Remembering that geodesic slicing is characterized by unit
  lapse and that for maximal slicing in \cite{mypaper1} the lapse at the
  right-hand event horizon has been found to approach \hbox{$3\sqrt{3}/16
  \approx 0.3248$} in the limit of late times, one can in fact observe that
  those values are obtained as pre-factors of the linear term. A deeper
  understanding of this observation would be an interesting topic for further
  investigation.}
\end{thebibliography}


\end{document}